\documentclass[nolinenumbers]{aastex631}

\submitjournal{PSJ}

\shorttitle{Barrel Instability in Binary Asteroids}
\shortauthors{\'Cuk et al.}
\graphicspath{{./}{figures/}}

\begin{document}

\title{Barrel Instability in Binary Asteroids}

\correspondingauthor{Matija \'Cuk}
\email{mcuk@seti.org}

\author[0000-0003-1226-7960]{Matija \'Cuk}
\affiliation{SETI Institute \\
189 N Bernardo Ave, Suite 200\\
Mountain View, CA 94043, USA}

\author[0000-0002-4952-9007]{Seth A. Jacobson}
\affiliation{Department of Earth and Environmental Sciences\\
Michigan State University\\
288 Farm Ln, Rm 207\\
East Lansing, MI 48824, USA}

\author[0000-0002-0906-1761]{Kevin J. Walsh}
\affiliation{Southwest Research Institute\\
1050 Walnut St. Suite 400\\
Boulder, CO 80302, USA}



\begin{abstract}
Most close-in planetary satellites are in synchronous rotation, which is usually the stable end-point of tidal despinning. Saturn's moon Hyperion is a notable exception by having a chaotic rotation. Hyperion's dynamical state is a consequence of its high eccentricity and its highly prolate shape \citep{wis84}. As many binary asteroids also have elongated secondaries, chaotic rotation is expected for moons in eccentric binaries \citep{cuk10}, and a minority of asteroidal secondaries may be in that state \citep{pra16}. The question of secondary rotation is also important for the action of the BYORP effect, which can quickly evolve orbits of synchronous (but not non-synchronous) secondaries \citep{cuk05}. Here we report results of a large set of short numerical simulations which indicate that, apart from synchronous and classic chaotic rotation, close-in irregularly-shaped asteroidal secondaries can occupy an additional, intermediate rotational state. In this ``barrel instability'' the secondary slowly rolls along its long axis, while the longest axis is staying largely aligned with the primary-secondary line. This behavior may be more difficult to detect through lightcurves than a fully chaotic rotation, but would likewise shut down BYORP. We show that the binary's eccentricity, separation measured in secondary's radii and the secondary's shape are all important for determining whether the system settles in synchronous rotation, chaotic tumbling, or barrel instability. We compare our results for synthetic asteroids with known binary pairs to determine which of these behaviors may be present in the Near-Earth Asteroid binary population.

\end{abstract}


\section{Introduction} \label{sec:intro}

There are many kinds of binary asteroids in the Solar System \citep{mar15, wal15}, but in this work we will focus on a relatively widespread and homogeneous sub-type. ``Typical small binary asteroids'' are a special class of binary asteroids that are common across multiple small-body populations, and consist of a km-sized rapidly rotating primary and a relatively large (with a radius ratio $R_2/R_1>0.2$) synchronously rotating and elongated secondary \citep{wal15}. Such systems are sometimes termed ``singly-synchronous binaries'', or (confusingly) ``asynchronous binaries'' to distinguish them from fully synchronous pairs where both components are spin-locked. 

Small binary asteroids are thought to be primarily formed by the spin-up of the primary to the critical rotation by the Yarkovsky-O'Keefe-Radzievskii-Paddack (YORP) effect \citep{rub00, bot06}, followed by the formation of the secondary which may be gradual \citep{wal08} or may happen in an episode of major instability \citep{jac11b}. As the secondaries tend to be irregularly shaped and are most commonly rotating synchronously, asymmetry between the re-emission of thermal radiation from their leading and trailing hemispheres (binary YORP, or BYORP, effect) should evolve their orbits on short ($<10^5$~yr for a typical NEA)  timescales \citep{cuk05, mcm10a}. Theorists have proposed rival pictures of small binary asteroids as fast-forming and fast-destroyed ephemeral systems ruled by radiation forces \citep{cuk07} or largely long-lived, stable systems in which tides and BYORP may be in equilibrium \citep{jac11a}. The first binary with observationally measured mutual orbital evolution, 1996~FG3, showed no orbital period change at all \citep{sch15}. This was widely seen as supporting evidence for tide-BYORP equilibrium.

More recent observational results \citep{sch19} for two other asteroids, show one (2001~SL9) having a rapidly shrinking orbit (presumably due to BYORP), while the other (Moshup) has a slowly expanding orbit (with ambiguous interpretation). It appears that not all small binaries are uniformly in equilibrium with zero orbital evolution. 

While BYORP explains fast orbital evolution, orbital migration seen in the other two systems that is much slower than the predicted BYORP drift \citep{mcm10b} requires a more complex theoretical explanation. A potential solution would be for the satellite of 1996~FG3 to be in asynchronous rotation, effectively turning off BYORP \citep{cuk10}. However, the lightcurve of 1996~FG3 does not show additional periods, and the long axis of the secondary seems to be consistently aligned with the primary \citep{sch15}. In addition, the orbital eccentricity of $e < 0.07$ may not be sufficient to induce chaotic rotation, according to the numerical results of \citet{cuk10}, despite being formally in the chaotic region \citep{wis84, nai15a}. However, the \citet{cuk10} model restricts the secondary's rotation to the two dimensions in the orbital plane, and as it is well-known that chaotic rotation changes character in three dimensions \citep{bla95, har11}, a more advanced model is clearly needed. 

In this paper, we present large-scale sets of integrations of the dynamics of synthetic binary asteroids. We vary numerous system parameters (size ratio, separation, eccentricity, primary and secondary shape) as well as the characteristics of system's heliocentric orbit (semimajor axis, eccentricity, obliquity between the system's plane and the heliocentric orbit). Our aim is to determine the combinations of parameters for which the secondary experiences non-synchronous rotation in a fully three-dimensional numerical model.


\section{Numerical Integrator}

In this work we use the numerical integrator {\sc br-sistem}, based on the integrator {\sc r-sistem} developed by \citet{cuk16b} for integrating the orbital and rotational evolution of Earth's Moon. {\sc br-sistem} ("Symplectic Integrator with Solar Tides in the Earth Moon system"; "B" stands for "binary" and "R" stands for "rotation") directly integrates both the orbital and rotational motion of the secondary. The satellite experiences tidal accelerations, gravity of the primary's equatorial bulge, and solar perturbations. The orbital part of the integrator is a symplectic mixed-variable integrator based on the principles of \citet{wis91}, with the specific implementation taken from \citet{cha02}. The orbital part of {\sc r-sistem} overlaps substantially with the more general-purpose satellite dynamics code {\sc simpl} \citep{cuk16a}. 

In {\sc br-sistem}, integration of a satellite's rotation is based on the Lie-Poisson approach of \citet{tou94}, with the secondary treated as a tri-axial rigid body torqued by the primary and the Sun. For internal dissipation, we adopted the approach similar to that of \citet{vok07}, where a torque perpendicular to the angular momentum vector pushes it toward the axis of the largest principal moment of inertia, with the torque's intensity adjusted to match the wobble damping timescales predicted by \citet{sha05}. The primary's axis is made to precess due to instantaneous torques by the secondary and the Sun on the equatorial bulge (i.e. we assume that the primary is purely oblate and in principal axis rotation). The tides within the primary and the secondary are calculated solely on the basis of instantaneous quantities such as positions and velocities, which are used to determine the position of the instantaneous tidal bulge which torques the other body. All tidal terms have been tested and the resulting damping closely follows expected analytical expressions (\citet{cuk16b} has a much more detailed description of, and the rationale for, our satellite tide model). For tidal parameters, we used the values from \citet{gol09}, i.e. Q=100 and $k_2=10^5 R$[km]. Note that the very short timescale of our integrations (100 yr) makes the details of the tidal model relatively moot, as we do not expect significant evolution over such short timescales.

\section{Design of the Dynamical Survey of Binaries}\label{sec: survey}

In order to determine the secondary spin state in a synthetic binary system, we have determined that a 100-year simulation is optimal as a part of a large survey. Since typical orbital periods are on the order of a day, 100 years gives us $10^4$ orbits to average the secondary's behavior. These integrations are also short enough that large numbers of them could be done in practicable amount of time (100-yr run in {\sc br-sistem} takes about a minute on a desktop CPU). 

In all our simulations, the mutual orbit is initially almost co-planar (within $0.01^{\circ}$) with the primary's equator. Primary's radius is set at 500~m, and the components' densities are 1500 kg m$^{-3}$. Primary's spin rate is $7 \times 10^{-4}$~rad/s, equivalent to a 0.1~day period. Since the primary is azimuthally symmetric and the synchronous orbit is well inside the secondary's orbit, the only dynamical relevance of primary rotation is for the primary's nutation and the system's precession. The secondary is initially placed at the pericenter of its orbit, and the semimajor axis and the eccentricity are converted to initial vectors using the geometric elements approach of \citet{ren06}. The secondary is initially assumed to be rotating synchronously with the (geometric element) mean motion, with the spin axis perpendicular to the orbit plane. The secondary is initially assumed to be rotating around its shortest axis, with the longest axis aligned with the primary-secondary line.

We are grouping our simulations in blocks of 2592 runs. Each run contains a five-dimensional array of initial conditions. Within a block we vary the primary's oblateness moment $J_2$ (0.02, 0.04), initial eccentricity (0, 0.02, 0.04, 0.06, 0.08, 0.1), secondary-to-primary (mean) radius ratio (0.2, 0.25, 0.3, 0.35, 0.4, 0.45), secondary medium-to-short (b/c) axis ratio (1.05, 1.1, 1.15, 1.2, 1.25, 1.3), and secondary's long-to-medium (a/b) axis ratio (1.1, 1.2, 1.3, 1.4, 1.5, 1.6). Mean radius of the secondary is  computed as volume equivalent radius $R_2=(a b c)^{1/3}$, and the mass is computed using a density of 1500~kg~m$^{-3}$. The range of values for each parameters is based on the parameters observed in the binary population.

Each block of 2592 runs is executed sequentially on a single processor, taking about two days of wall time. These blocks were combined into groups of twenty to make simulations sets. All the blocks of a set were run parallel on twenty different CPUs of a multi-processor workstation. Within each set, we varied obliquity of the system relative to the heliocentric orbit (0-180$^{\circ}$, five values) and the initial mutual semimajor axis (four values). The ranges of initial mutual semimajor axis varied between sets, as did heliocentric semimajor axis and eccentricity (Table \ref{table1}). Overall, we ran four sets of 20$\times$2592=51,840 simulations, equaling a total of 207,360 simulations. The decision on the number and parameters of simulation sets was based on preliminary results, which prompted us to explore a wider range of mutual semimajor axis than initially planned, whereas we decided against a wide exploration of heliocentric orbital elements as their apparent lack of importance did not justify further use of computational resources.

\begin{table}
\centering
\caption{Parameters of the system in our four sets  of simulation blocks.\label{table1}}
\begin{tabular}{lcccc}
\hline
Designation & Heliocentric $a$ & Heliocentric $e$ & System obliquity $\theta$ & Mutual $a$ [$R_1$]  \\
\hline
a111030 & 1.1~AU & 0.3 & $0^{\circ}$, $45^{\circ}$, $90^{\circ}$, $135^{\circ}$, $180^{\circ}$ &3, 4 5, 6  \\
a116055 & 1.6~AU & 0.55 & $0^{\circ}$, $45^{\circ}$, $90^{\circ}$, $135^{\circ}$, $180^{\circ}$ &3, 4, 5, 6 \\
a211030 & 1.1~AU & 0.3 & $0^{\circ}$, $45^{\circ}$, $90^{\circ}$, $135^{\circ}$, $180^{\circ}$ &7, 8, 9, 10  \\
a216055 & 1.6~AU & 0.55 & $0^{\circ}$, $45^{\circ}$, $90^{\circ}$, $135^{\circ}$, $180^{\circ}$ &7, 8, 9, 10 \\
\hline
\end{tabular}
\end{table}

\section{Dynamical Behaviors in Numerical Simulations}\label{sec:barrel}

In order to classify the dynamical states of the secondary, in this section we will restrict ourselves to one block of the a111030 set of simulations. This block (designated ``o1a1'') had the initial obliquity of the system set to $180^{\circ}$ and the initial semimajor axis was set to three primary radii (i.e. $a=1500$~m). Within the 2592 simulations we varied primary oblateness, initial eccentricity, component radius ratio, and the secondary's triaxial shape as described in the previous section. 

\begin{figure}
\epsscale{.6}
\plotone{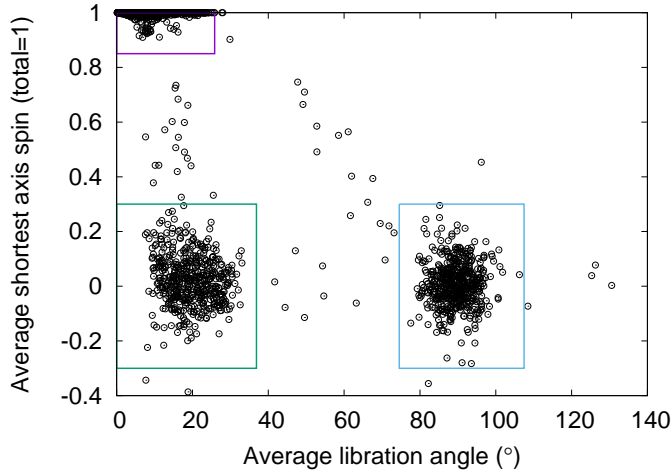}
\caption{The average short-axis angular momentum (normalized by the total angular momentum) against the secondary's average libration angle for the secondary in a block of 2592 simulations. This block of the simulation set a111030 has system obliquity set to $180^{\circ}$ and mutual separation to $a=3 R_1$. The average libration angle was obtained by taking the mean of the value $\cos{\psi}$ over the course of the simulation, with $\psi$ being the angle between the primary and the secondary's long axis. Bodies in principal axis rotation plot at the top of the figure, while bodies in synchronous lock (i.e. with long axis aligned with the primary-secondary line) plot at the left edge of the figure. Three groupings are immediately clear: synchronous secondaries (in the magenta box), chaotic secondaries (light blue box) and the secondaries affected by the so-called ``barrel instability'' (dark green box).}
\label{fig1}
\end{figure}

Figure \ref{fig1} shows the average short-axis spin (i.e. angular momentum) of the secondary (normalized to the total) against the secondary's average libration angle. The simulation largely cluster into three groups, which we define using boxes in Fig \ref{fig1}. Simulations in the upper left corner of Fig. \ref{fig1} have the secondary in synchronous principal-axis rotation. The cluster around zero average short-axis spin and $90^{\circ}$ average libration angle (i.e. $\left<\cos{\psi}\right>=0$) represents secondaries in chaotic rotation. For them, there is no preferred short-axis spin direction and no preferred angle between the long axis and the primary-secondary line. The third cluster represents a new dynamical state, in which the rotation is close to chaotic, but the secondary keeps the long axis approximately aligned with the primary. In this state the secondary does not switch the nearside and the farside hemispheres, but it does switch the north and south hemispheres through rotation along the long axis. We term this state the ``barrel instability'', as the secondary is periodically rolling along its longest axis, somewhat like a barrel. The points outside the boxes around the main clusters are classified as ``other'', and they typically switched from one state to another during the simulation, so their average parameters are not representative of a long-term state.

\begin{figure}
\epsscale{1.2}
\plotone{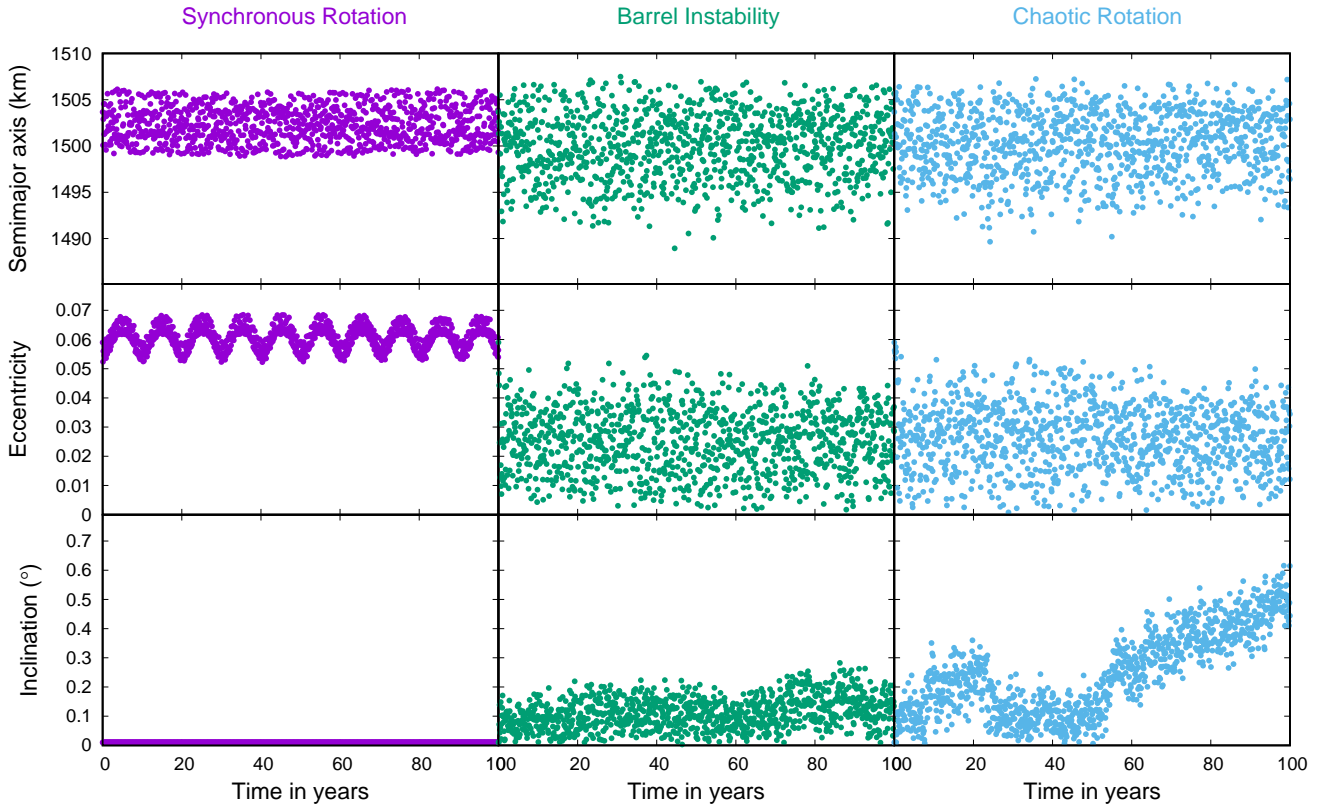}
\caption{Comparison of the orbital elements of three simulations identical in their initial conditions, except for the secondary's elongation (see text). The three simulations resulted in three different dynamical behaviors: synchronous rotation (left-hand-side panels), barrel instability (middle panels), and chaotic rotation (right-hand-side panels). Based on orbital parameters alone, barrel instability is hard to distinguish from a fully chaotic rotation. For the same initial eccentricity the non-synchronous simulations have a lower long-term eccentricity as angular momentum deficit is being absorbed by the secondary's rotation. Inclinations stay below one degree over the course of the simulation for all three simulations.}
\label{orbit}
\end{figure}

To illustrate the characteristics of these three kinds of dynamical behavior, we selected three simulations from the block plotted in Fig.~\ref{fig1} for closer inspection. We chose three simulations identical in all initial conditions except the secondary's elongation, but each simulation demonstrated a different example of the three dynamical behaviors. All three simulations had primary $J_2=0.02$, initial eccentricity $e_0=0.06$, radius ratio $R_2/R_1=0.25$, and secondary medium-to-short axis ratio of $b/c=1.25$. The secondary's elongation was $a/b=1.2$ for the simulation with synchronous rotation, $a/b=1.3$ for the run with barrel instability, and $a/b=1.1$ for the case with chaotic rotation. In Fig.~\ref{orbit} we have plotted side-by-side the binary's semimajor axis, eccentricity, and inclination (measured to primary's equator) over the course of the three 100-yr simulations. The most interesting feature of their orbital evolution is the {\it lower} average eccentricity of the two simulations with the excited secondary rotation. The reason for this is that the angular momentum deficit (AMD) of the orbit, defined as $\sqrt{GM a}(1-\sqrt{1-e^2} \cos{i})$ \citep{las97}, is apparently absorbed by the excited rotation of the secondary, leaving the orbit more circular. 

\begin{figure}
\epsscale{1.2}
\plotone{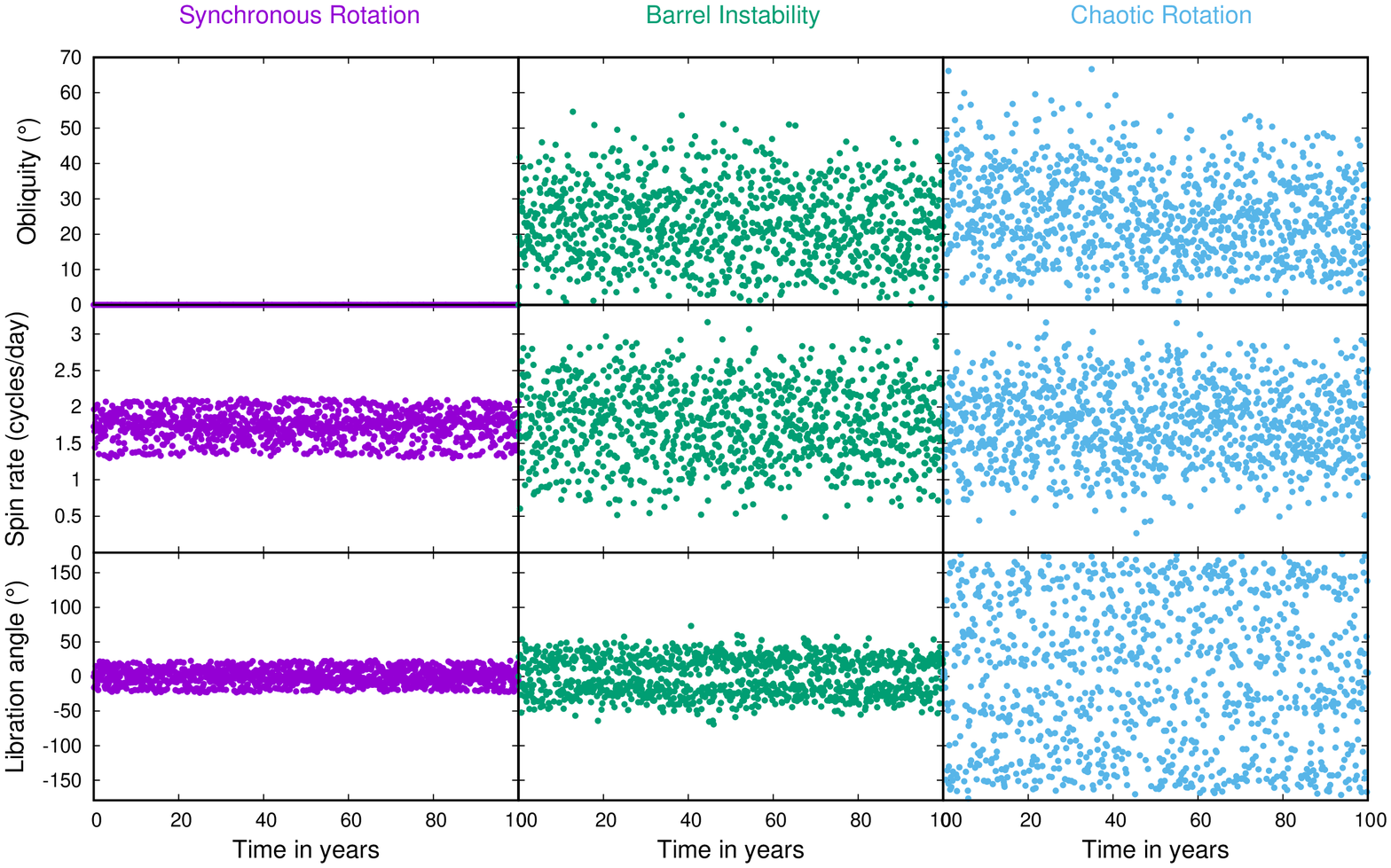}
\caption{Comparison of the secondary's rotational parameters in the three simulations plotted in Fig.~\ref{orbit}. While both the barrel instability and chaotic rotation simulations exhibit wide swings in the secondary's obliquity (the angle between the instantaneous spin vector and the mutual orbital plane) and spin rate, the barrel instability differs from full chaotic rotation because the satellite's long axis is mostly aligned with the primary (bottom center panel).}
\label{spin}
\end{figure}

\begin{figure}
\epsscale{1.2}
\plotone{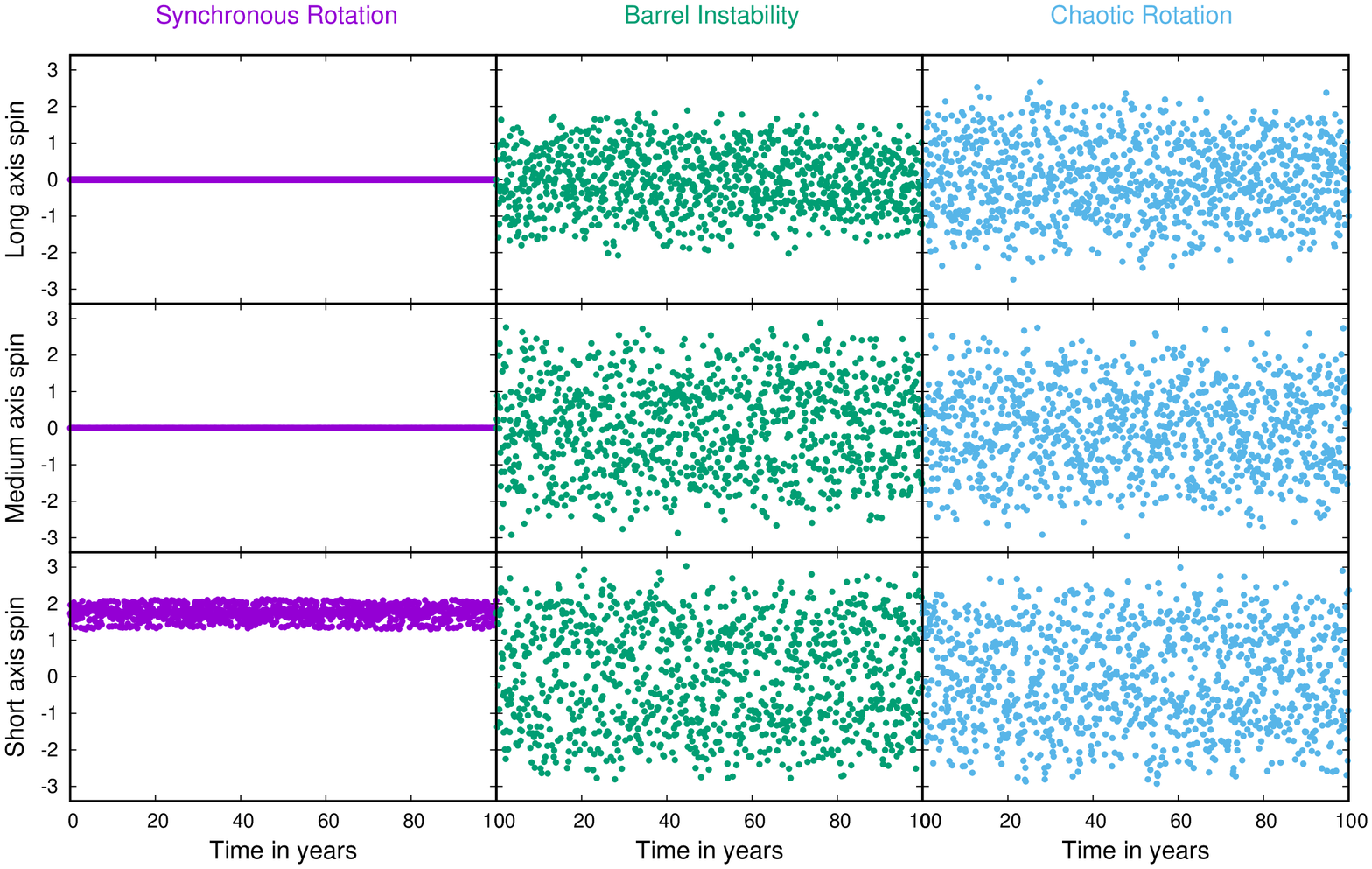}
\caption{Comparison of the secondary's rotation rates around its three principal axes (in the units of cycles per day) in the three simulations plotted in Fig.~\ref{orbit}. While the system undergoing the barrel instability exhibits a different behavior in the orbital reference frame, the difference between this dynamical mode and a fully chaotic rotation is not pronounced in the body frame.}
\label{body}
\end{figure}

Unlike the orbital parameters, the spin parameters of the secondaries uniquely identify the dynamical behavior of the same three simulations, as shown in Fig.~\ref{spin}. Unlike the synchronous simulation in which the spin of the secondary is both restricted to the orbital plane and close to the synchronous rate, the two excited simulations exhibit a large range of obliquities (up to $70^{\circ}$). Likewise, the spin rates of both excited simulations are varying over a wider range of values than the one in the synchronous run. The libration angle plots (bottom row in Fig.~\ref{spin}) show the only unambiguous difference between the barrel instability and a fully chaotic rotation. The system experiencing barrel instability (bottom center panel in Fig.~\ref{spin}) has the secondary's longest axis staying within about $50^{\circ}$ of the direction to the primary (this includes both in-plane and out-of-plane misalignment). On the other hand, spin frequencies around the secondary's three principal axes (Fig.~\ref{body}) are very similar between the barrel instability and chaotic rotation simulations, with the only discernible difference being the narrower range of spin rates around the longest axis seen for the barrel instability.

\begin{figure}
\epsscale{1.}
\plotone{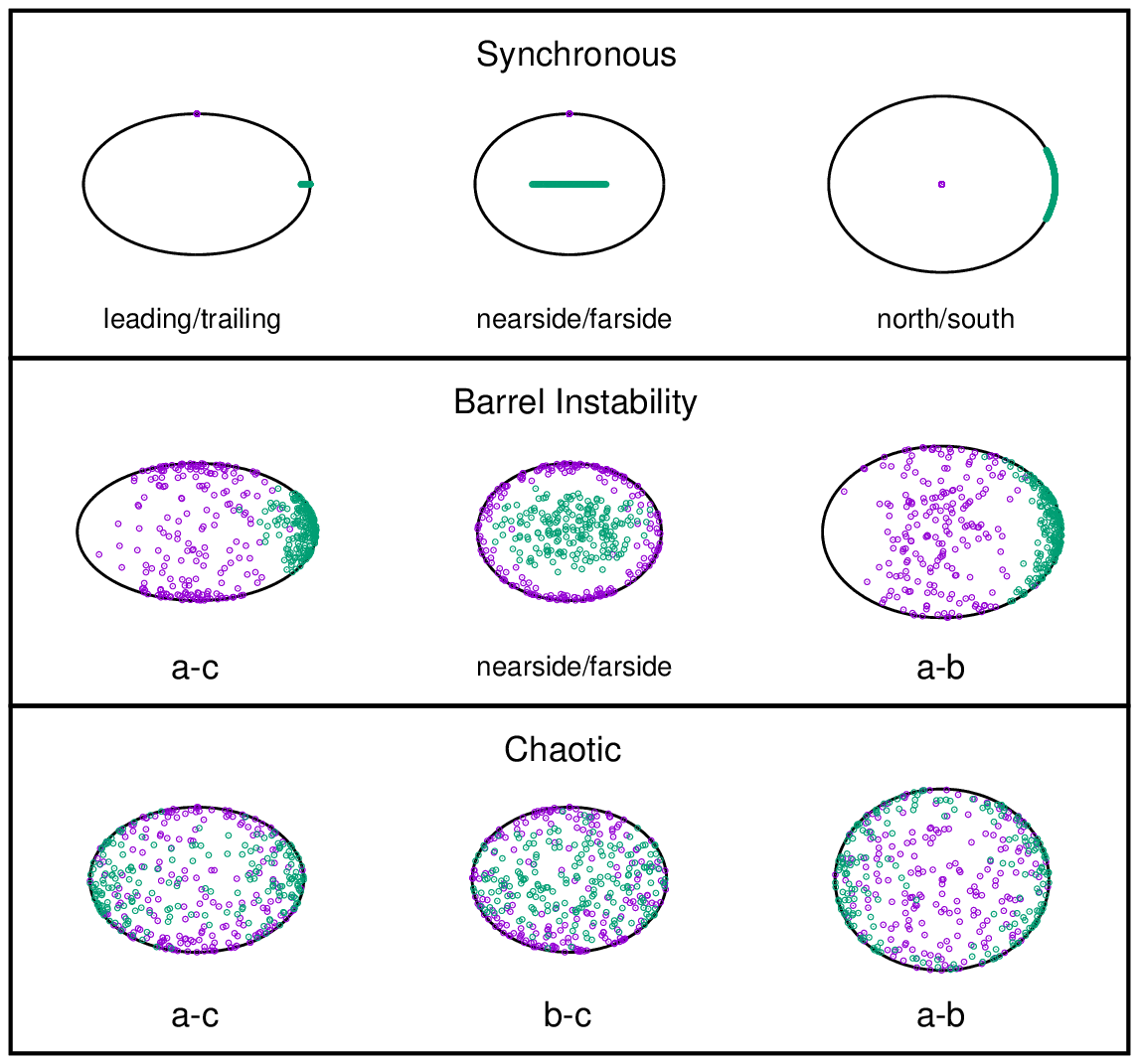}
\caption{Instantaneous rotation poles (magenta points) and sub-primary points (dark green) on the surfaces of the secondaries in the three simulations plotted in Figs. \ref{orbit}-\ref{body}. The secondaries are transparent, and the three views are along the three principal axes (left to right): intermediate, longest and shortest. In the case of the synchronous secondary, these views can be considered to show the leading and trailing sides, nearside and farside, and the north and south hemispheres (in each case both sides are visible because of transparency). In the case of barrel instability, nearside and farside labels are still meaningful, but the body does not have permanent leading and trailing or north and south hemispheres.}
\label{globes}
\end{figure}

In order to illustrate the nature of barrel instability in the secondary's reference frame, in Fig. \ref{globes} we plot the locations of the instantaneous rotation pole and the sub-primary point on the surfaces of the secondary in the three simulations shown in Figs. \ref{orbit}--\ref{body}. While ``barrel instability'' has not been discussed in detail in the binary asteroid literature, it has been observed in full two-body dynamics simulations using the {\sc gubas} (General Use Binary Asteroid Simulator) code \citep{dav20, mey21, agr21}. Furthermore, this kind of rotational behavior has been seen in studies of spacecraft dynamics in the presence of a gravity gradient \citep{hug86, sch09}. 

\begin{figure}
\epsscale{.6}
\plotone{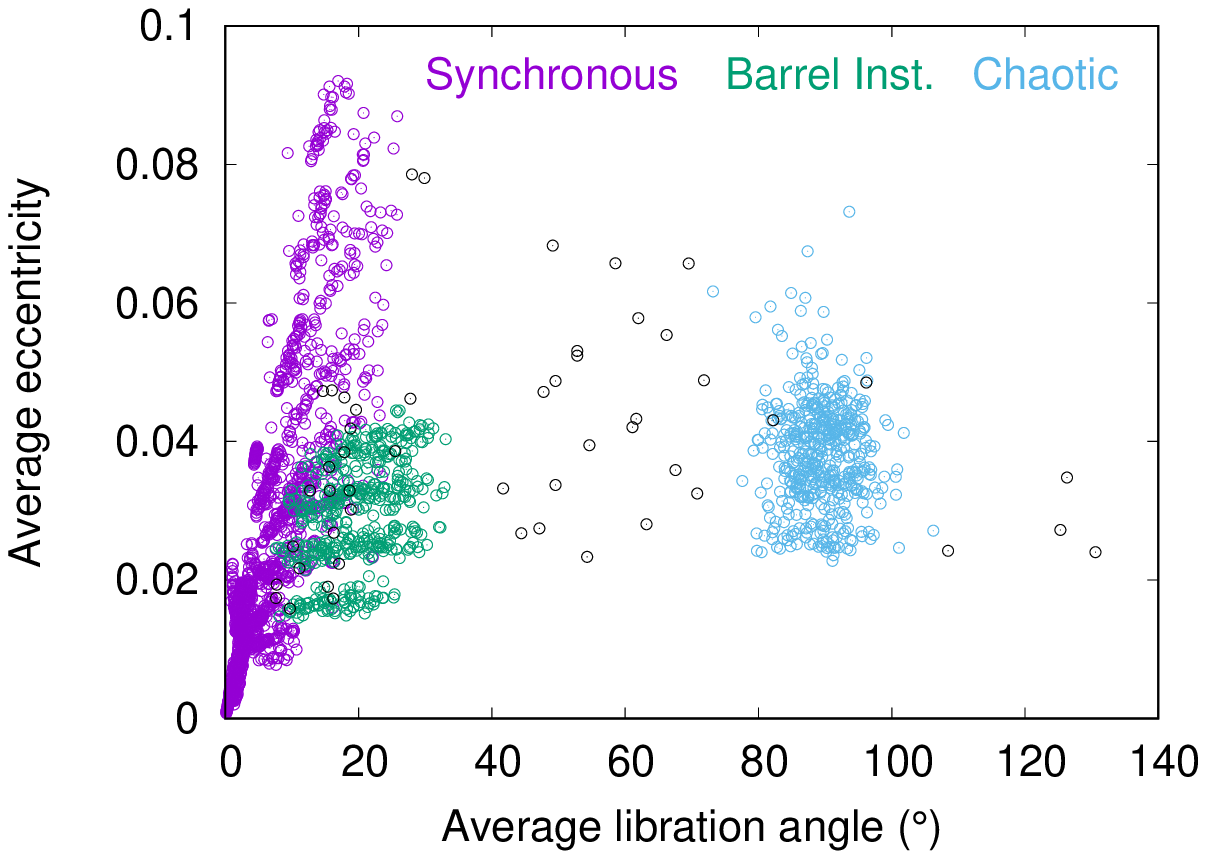}
\caption{The average eccentricity potted against the secondary's average libration angle for the secondary in the same block of 2592 simulations shown in Fig.~\ref{fig1}. The three groupings are plotted in different colors: synchronous secondaries are plotted in magenta, chaotic secondaries in light blue, and the secondaries affected by the barrel instability are plotted in dark green (non-clustered points are shown in black). While the lowest average eccentricity systems are synchronous (including all with the initial $e_0 \le 0.02$), excited systems, especially those undergoing barrel instability, can have lower long-term eccentricities than many of the synchronous systems.}
\label{ecc}
\end{figure}

To emphasize that eccentricity is not an effective diagnostic for dynamical behavior, Fig.~\ref{ecc} shows the same simulation block as Fig.~\ref{fig1} but instead plots average eccentricities of the mutual obit versus the average secondary libration angle. The three dynamical groupings are color-coded consistently with the boxes defining them in Fig.~\ref{fig1}. It is clear from Fig.~\ref{ecc} that there is no simple way of identifying the dynamical state of the system from its observed eccentricity. While some of the lowest-eccentricity systems are synchronous, so are the most eccentric ones, while the pairs with dynamically excited secondary typically fall in the middle.

\section{Dependence of Binary Dynamics on System Parameters}\label{sec:dependence}

\begin{figure}
\epsscale{1.}
\plotone{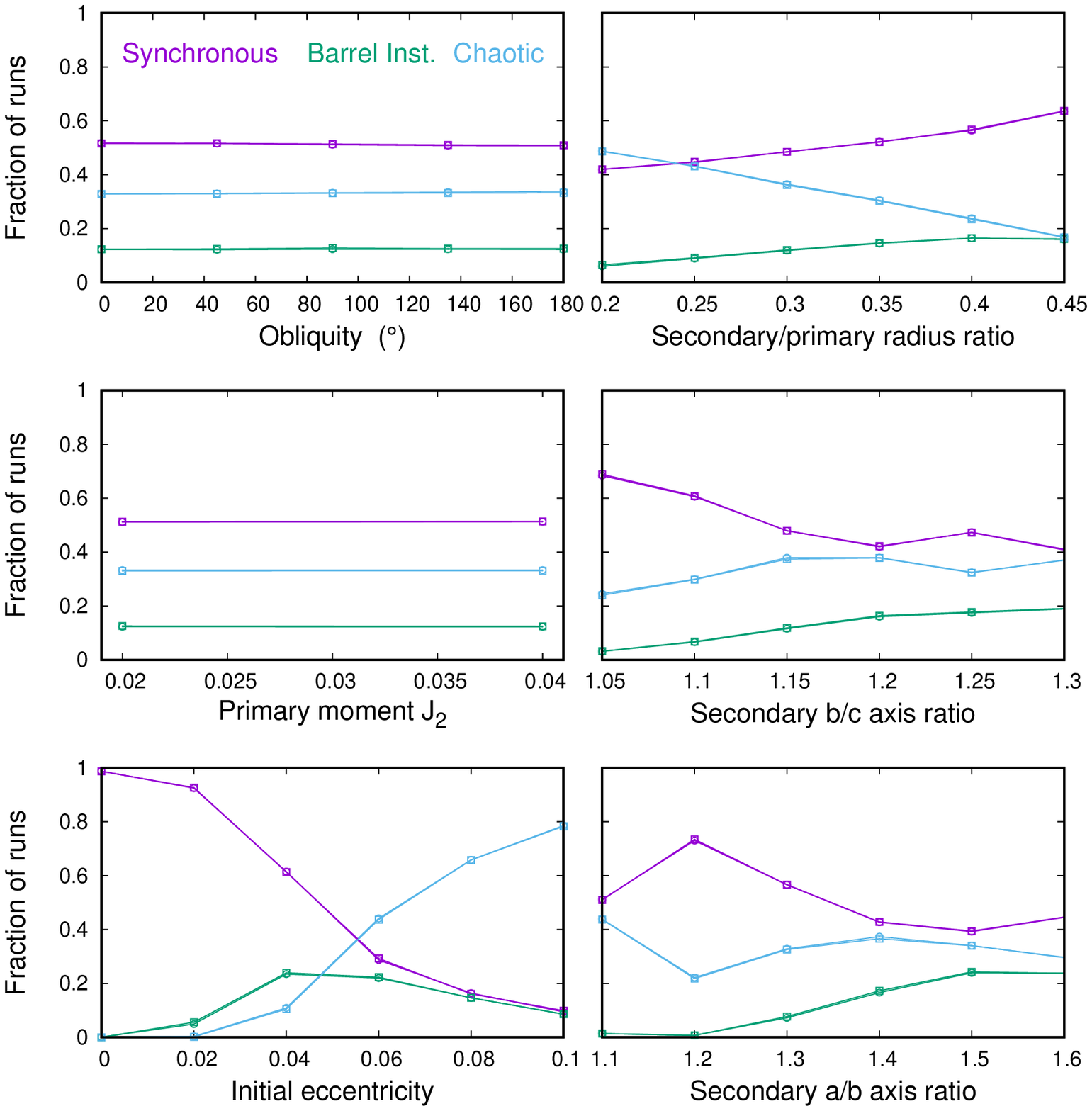}
\caption{Dependence of the dynamical outcomes on the initial conditions of our binary systems. Shown are outcomes in the simulation sets a111030 (circles) and a116055 (squares) that had $3 \le a \le 6$ $R_1$ (Table \ref{table1}). The outcomes for the two sets are practically indistinguishable, indicating that the heliocentric orbits has no major effect on dynamics of close binaries. The lack of dependence of dynamical outcome on the system's obliquity (top left panel) reinforces this point. Primary oblateness also appears to be of little importance for the values we explored. The dynamical outcome appears most dependent on the initial eccentricity, as well as the secondary's size and shape.}
\label{corr1}
\end{figure}

\begin{figure}
\epsscale{1.}
\plotone{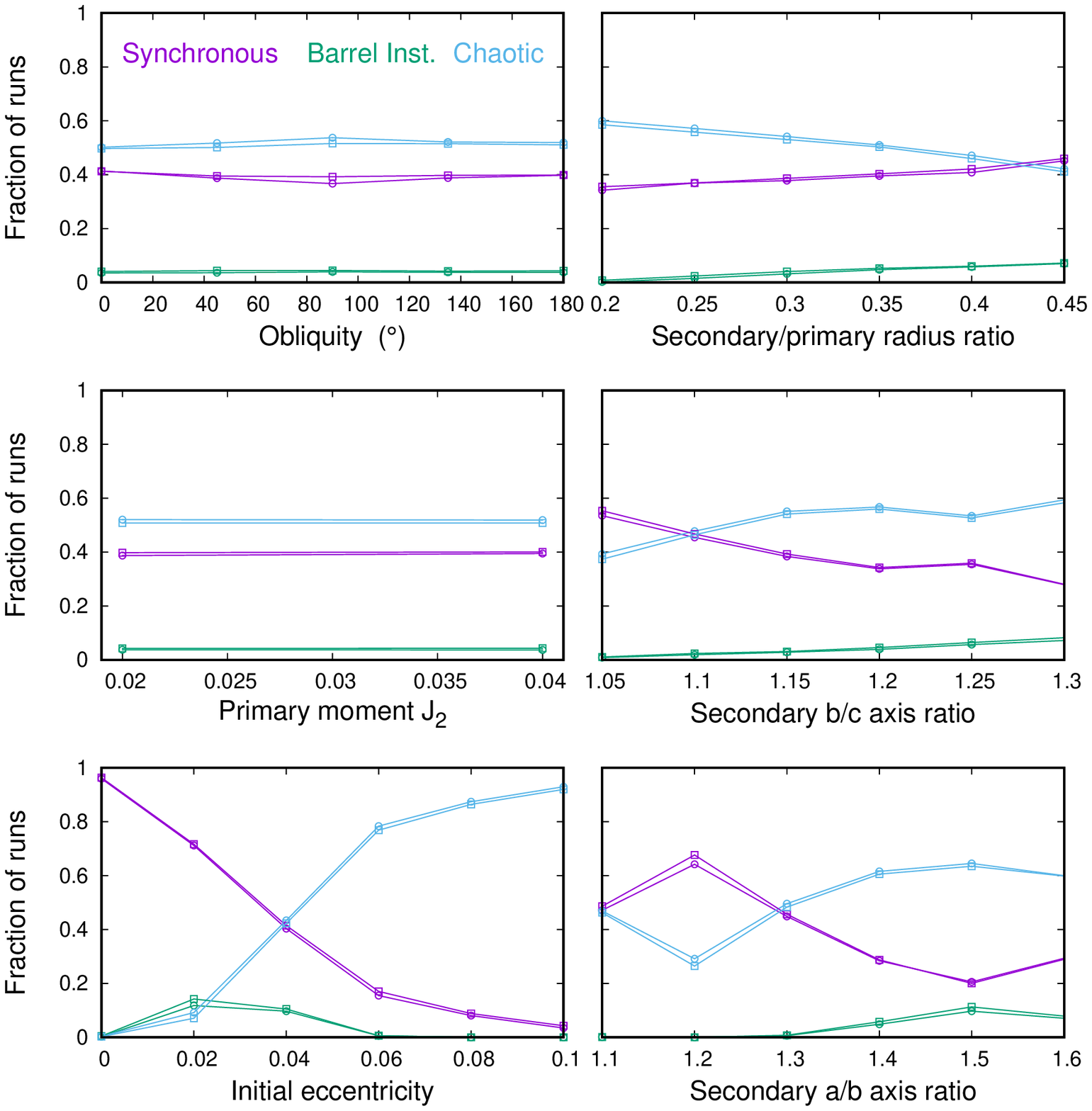}
\caption{Dependence of dynamical outcomes on initial conditions in simulation sets a211030 (circles) and a216055 (squares) that had $7 \le a \le 10$ $R_1$ (Table \ref{table1}). Unlike in Fig. \ref{corr1} that plots results for tighter binaries, these more separated pairs do experience some variation of outcome due to solar perturbations. Solar influence appears most noticeable for obliquities of $90^{\circ}$ and largest separations (Fig. \ref{all_a}), suggesting that the binaries are being perturbed through the Kozai-Lidov mechanism \citep{nao13}}
\label{corr2}
\end{figure}

\begin{figure}
\epsscale{.6}
\plotone{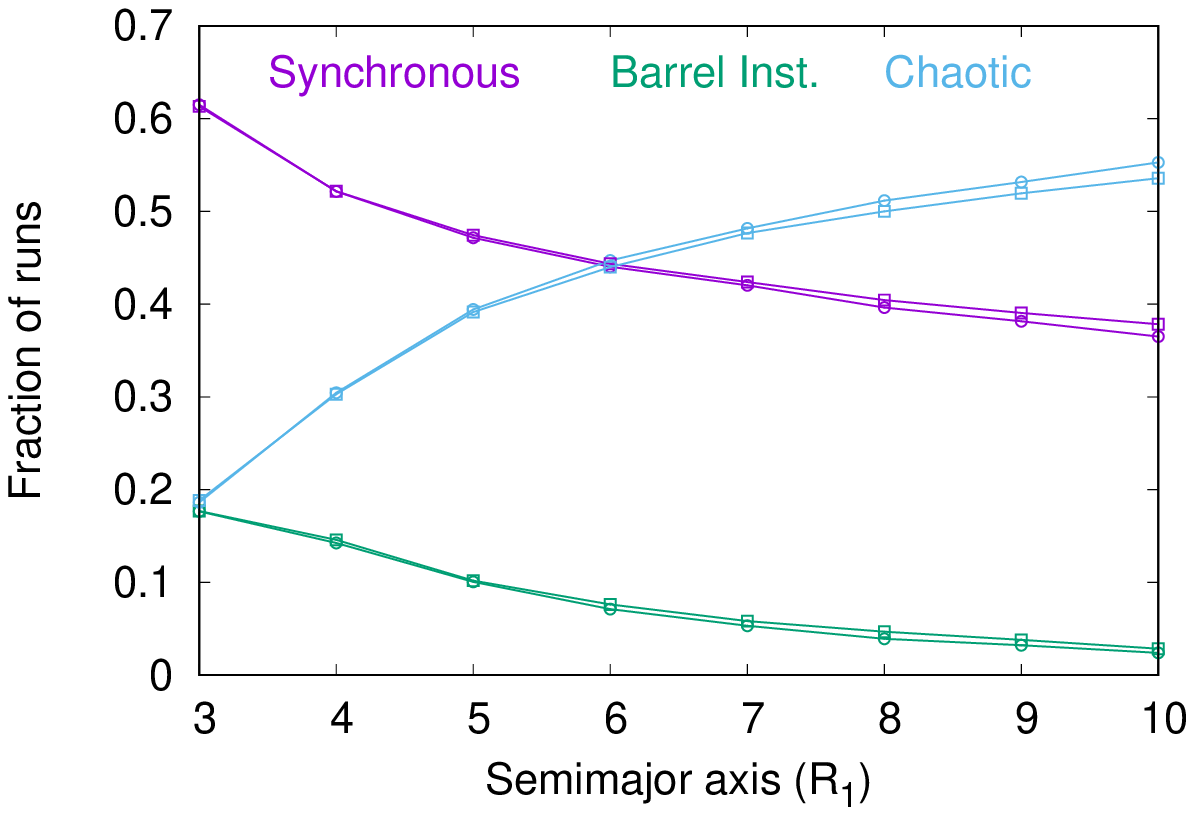}
\caption{Dependence of dynamical outcome on the initial semimajor axis of the binary across the whole dynamical survey. Results for simulation sets a111030 and a211030 are plotted by circles, and those for sets a116055 and a216055 are plotted by squares.}
\label{all_a}
\end{figure}

After having established the main classes of dynamical behavior for our simulated binary systems, we now turn our attention to the whole sample of 207,360 simulations. To determine which parameters are relevant for the secondary's dynamical behavior, within each of the four sets of simulation (Table \ref{table1}) we have created a matrix which kept track of dynamical outcomes (synchronous, barrel instability, chaotic, or other) for each data point in seven-dimensional space of our initial parameters: primary's oblateness moment $J_2$, initial eccentricity $e_0$, secondary-to-primary (mean) radius ratio $R_2/R_1$, secondary's medium-to-short ($b/c$) axis ratio, secondary's long-to-medium ($a/b$) axis ratio, obliquity of the system relative to the heliocentric orbit, and the initial mutual semi-major axis $a_0$. From this matrix we were able to generate plots of the numbers of different outcomes as a function of each parameter. Such plots for six of our initial parameters are plotted in Figs.~\ref{corr1} and \ref{corr2}; Fig.~\ref{corr1} plots results for the sets with $3 \leq a \leq 6~R_1$, while Fig.~\ref{corr2} plots results for the other two sets in which $7 \leq a \leq 10~R_1$. Dependence of outcomes on the mutual semimajor axis across all four sets is plotted in Fig. \ref{all_a}.

Figure \ref{corr1} indicates that the parameters of the heliocentric orbit have little effect on the dynamics of close binaries. While we have explored only two different heliocentric orbits, the fact that there is no change of outcome with changing obliquity strongly suggests that the contribution from solar perturbations is small. Semi-secular resonances with the Sun \citep{tou98} have been proposed as relevant for binary asteroid dynamics \citep{mcm16}, but such resonances should affect prograde and retrograde systems in fundamentally different ways, which is not observed. Similarly, there appears to be no change in the rotational state of the secondary when the primary's oblateness moment $J_2$ changes between $0.02$ and $0.04$. While larger $J_2$ may lead to changes in dynamics, we consider larger values to be unlikely for spheroidal (rather than highly irregular or contact-binary) primaries typical of small binary asteroids.

Figure \ref{corr2} plots the same quantities as Fig. \ref{corr1}, but for the sets of simulations with more separated binaries. Unlike in Fig. \ref{corr1}, here we see some variation of outcome between systems with different heliocentric orbital elements. In particular, it appears that the systems with a smaller heliocentric semimajor axis have a relative over-abundance of chaotic secondaries (and a deficit of synchronous ones) compared to systems with the greater heliocentric distance. This variation due to the heliocentric orbit is most noticeable for systems with obliquities of $90^{\circ}$ (Fig \ref{corr2}, top left panel) and larger separations (Fig. \ref{all_a}), indicating that Kozai-Lidov-type perturbations of the Sun \citep{nao13} are inducing additional eccentricity for high-obliquity systems. 

Figure \ref{all_a} combines all four sets of simulations to plot the dependence of the secondary's dynamics on the initial semimajor axis of the binary. Overall, it is clear that synchronous rotation dominates for the systems in our survey with smaller separations, while chaotic rotation becomes more widespread for larger separations. Barrel instability is less widespread than the other two outcomes, and is most common at relatively small separations. 

Comparison between Figure \ref{all_a} and top-right panels in Figs. \ref{corr1} and \ref{corr2} shows great similarities. Number of systems with chaotic rotation increases and those with synchronous rotation deceases as the mutual separation is increased, or when the size of the secondary is decreased. We suspect that both of these trends are a reflection of an importance of a single parameter: binary separation in terms of the secondary radius. As long as the secondary is much less massive than the primary, and the spin-orbit interactions of the primary are suppressed (due to fast rotation and spheroidal shape), angular momentum is mostly passed between the binary orbit and the secondary's rotation. As the parameter $a/R_2$ determines the relative sizes of the secondary's orbital and spin angular momenta (for synchronous rotation $L_{spin}/L_{orbit} \simeq (a / R_2)^{-2}$), it can be expected that it would be crucial for the systems dynamics.

\begin{figure}
\epsscale{.6}
\plotone{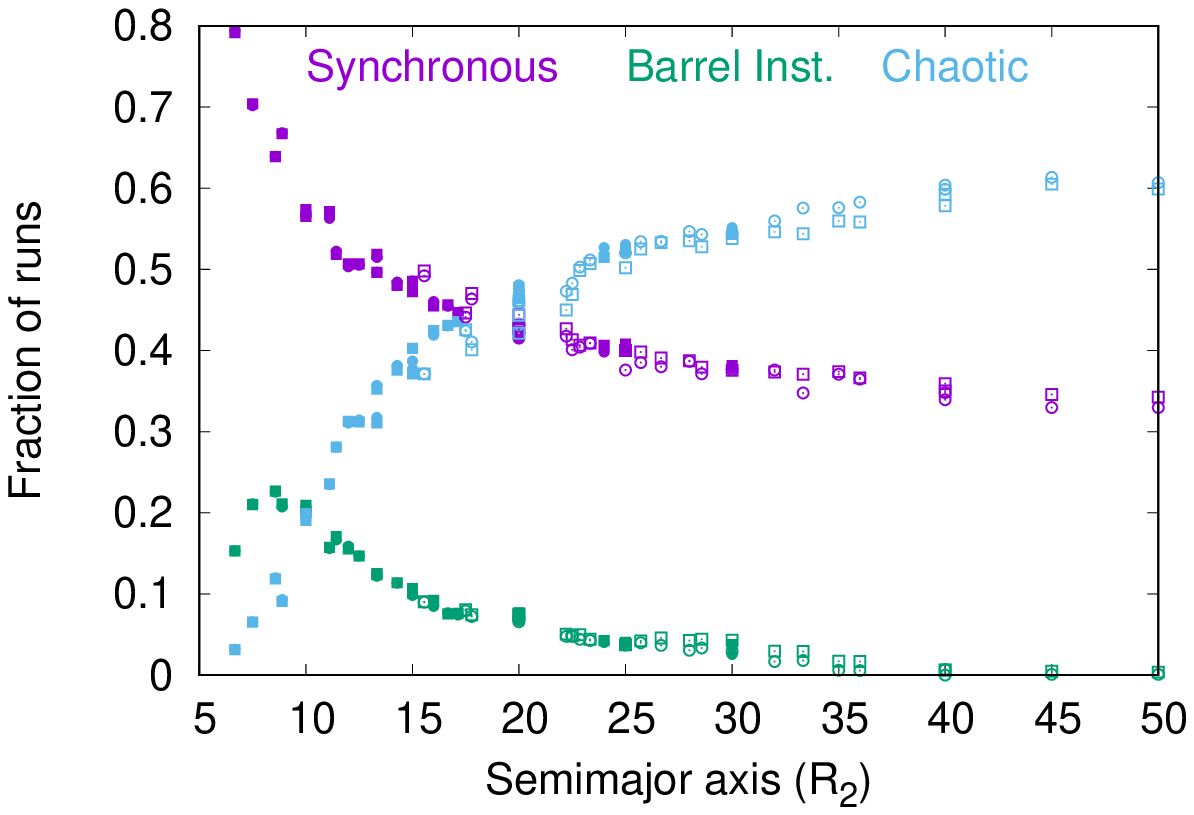}
\caption{Dependence of dynamical outcome on the initial semimajor axis of the binary expressed in secondary radii across the whole dynamical survey. Results for simulation sets a111030 and a211030 are plotted by solid and open circles, respectively, and those for sets a116055 and a216055 are plotted by solid and open squares.}
\label{all_r2}
\end{figure}

Fig. \ref{all_r2} re-plots the dynamical outcomes for all our runs as a function of initial binary semimajor axis in terms of the secondary radius. As the initial conditions were a grid in $a/R_1$ and $R_2/R_1$, there are now sometimes multiple data points for identical (or very similar) values of $a/R_2$. Nevertheless, the prevalence of our dynamical outcomes clearly clusters around well-defined lines, indicating that $a/R_2$ is a fundamental property of the system. The variation in the prevalence of different dynamical outcomes in Fig. \ref{all_r2} is more pronounced than in Fig. \ref{all_a} and top-right panels of Figs. \ref{corr1} and \ref{corr2}, suggesting that $a/R_2$ is actually a property of the system more fundamental for determining dynamics than $a/R_1$ and $R_1/R_2$, in line with our theoretical expectations.  

\begin{figure}
\epsscale{.6}
\plotone{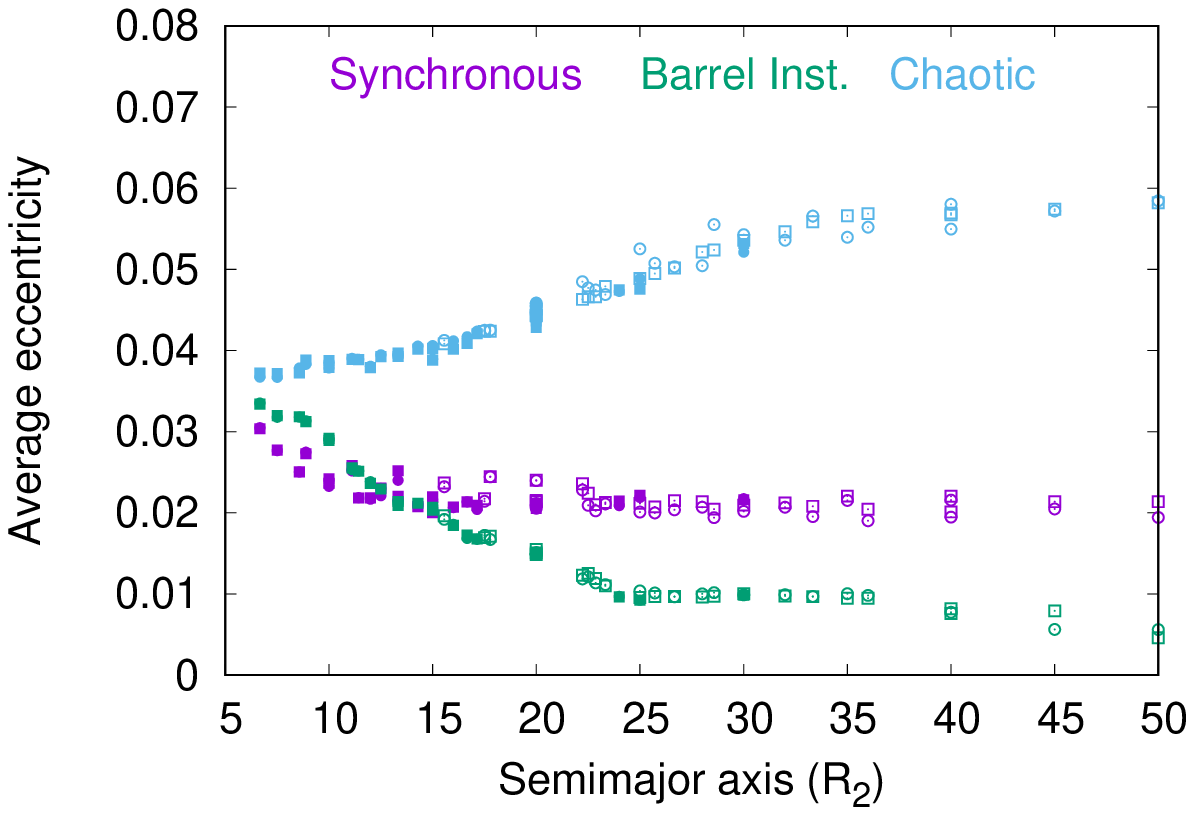}
\caption{Dependence of the average binary eccentricity in a simulation on the initial semimajor axis of the binary expressed in secondary radii across the whole dynamical survey. Results for simulation sets a111030 and a211030 are plotted by solid and open circles, respectively, and those for sets a116055 and a216055 are plotted by solid and open squares.}
\label{all_e2}
\end{figure}

Apart from $a/R_2$, initial eccentricity is clearly an important parameter for determining the system's dynamics. Within our $0 \le e_0 \le 0.1$ range, low initial eccentricities lead to synchronous rotation, high initial eccentricities lead to chaotic rotation, and barrel instability is most prevalent at medium values. Note that initial eccentricity is not an observable quantity, as it may immediately be changed by secondary's non-synchronous motion (Fig.~\ref{orbit}). Figure \ref{all_e2} plots average eccentricities for every dynamical outcome as a function of separation in secondary radii ($a/R_2$). At larger separations, chaotically rotating secondaries tend to be more eccentric, while synchronous systems have lower eccentricities (with the few systems with barrel instability being on average the least eccentric). At smaller separations ($a/R_2 \le 10$) the samples of the three major dynamical behaviors have more or less the same average eccentricity, making the observed eccentricity by itself not dynamically diagnostic for close-in systems.

\begin{figure}
\epsscale{.6}
\plotone{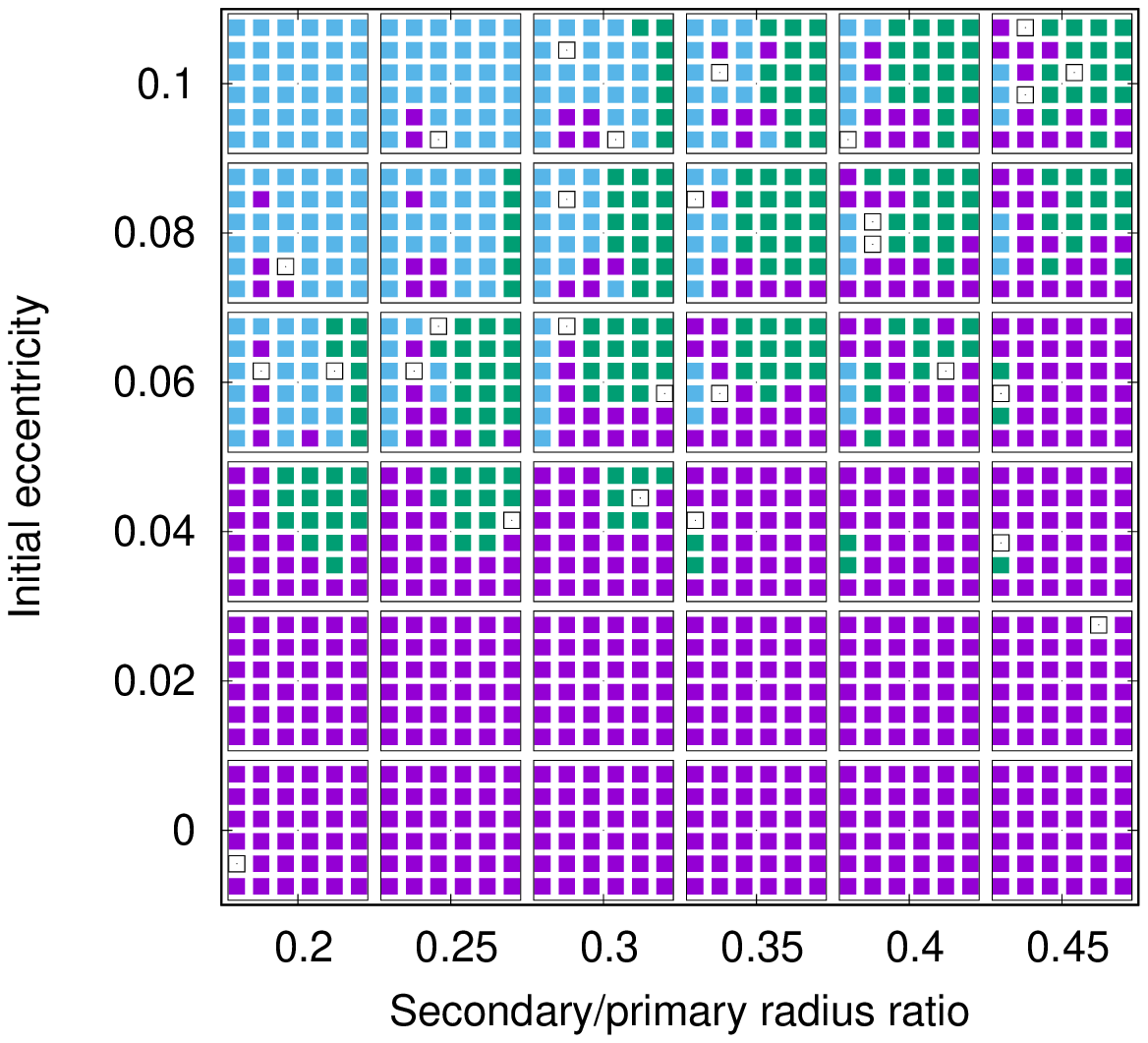}
\caption{Dependence of the dynamical outcome on the parameters of the binary. Each square point represents a simulation, and magenta symbols indicate synchronous rotation, dark green symbols indicate barrel instability, light blue symbols indicate chaotic rotation, and open symbols plot other (unclassified) outcomes. Each small 6x6 square represents dependence on the secondary's shape: the axis ratio $a/b$ changes from 1.1 to 1.6 left to right, and the axis ratio $b/c$ changes from 1.05 to 1.3 bottom to top. These small squares are then arranged according to the relevant secondary/primary radius ratio (on x axis) and initial eccentricity (on y axis). These simulations had primary oblateness $J_2$=0.02, binary separation $a=3 R_1$, obliquity $\theta=180^{\circ}$, and heliocentric elements $a_H$=1.1~AU and $e_H=0.3$.}
\label{tiles1}
\end{figure}

\begin{figure}
\epsscale{.6}
\plotone{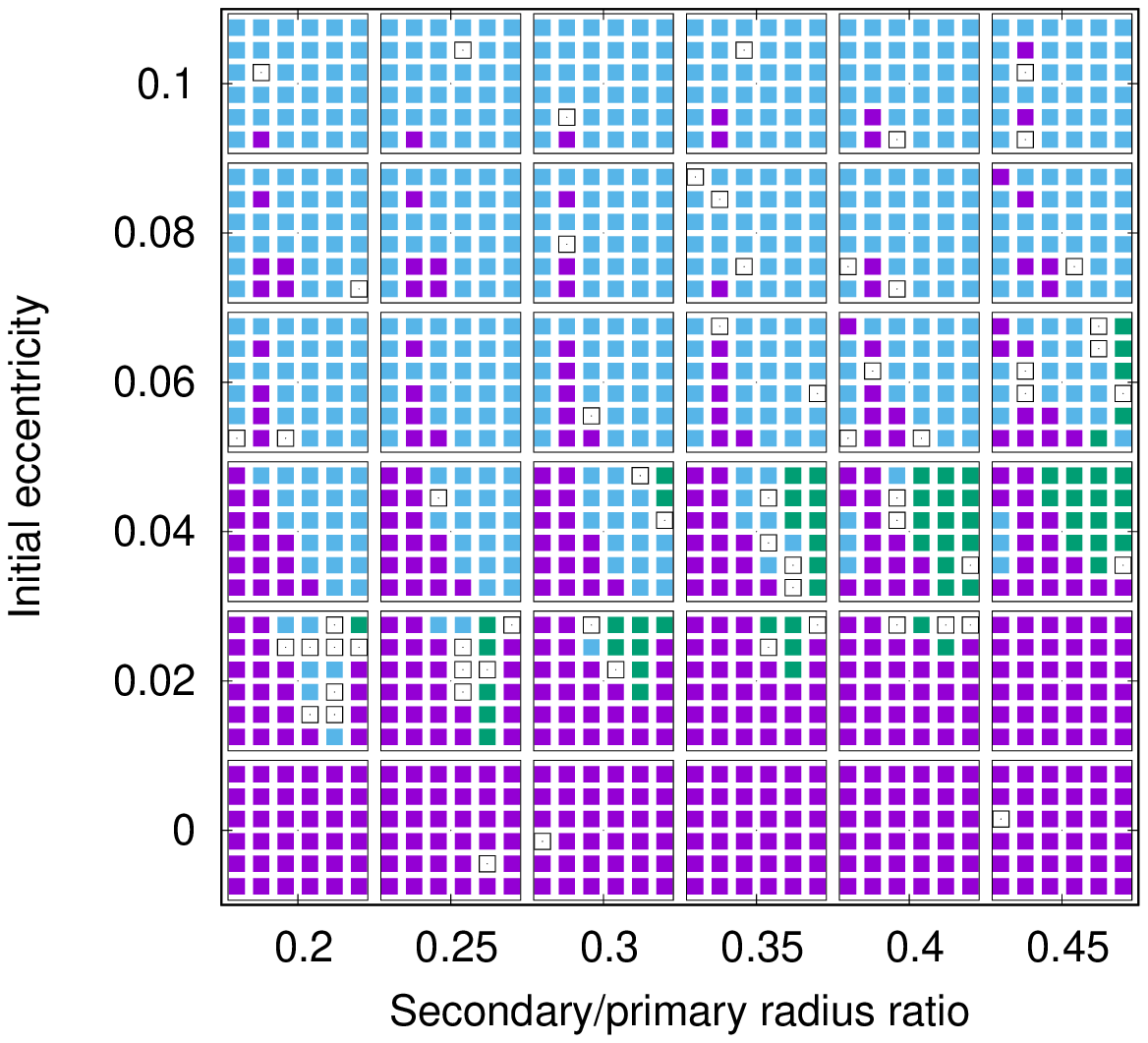}
\caption{Dependence of the dynamical outcome on the parameters of the binary. Each square point represents a simulation, and magenta symbols indicate synchronous rotation, dark green symbols indicate barrel instability, light blue symbols indicate chaotic rotation, and open symbols plot other (unclassified) outcomes. Each small 6x6 square represents dependence on the secondary's shape: the axis ratio $a/b$ changes from 1.1 to 1.6 left to right, and the axis ratio $b/c$ changes from 1.05 to 1.3 bottom to top. These small squares are then arranged according to the relevant secondary/primary radius ratio (on x axis) and initial eccentricity (on y axis). These simulations had primary oblateness  $J_2$=0.02, binary separation $a=7 R_1$, obliquity $\theta=180^{\circ}$, and heliocentric orbital parameters $a_H$=1.1~AU and $e_H=0.3$.}
\label{tiles2}
\end{figure}

Beyond the separation to secondary size ratio and initial eccentricity, the dynamical outcome depends in a non-trivial way on the secondary's axis ratios. In our survey, we varied both the $b/c$ and $a/b$ axis ratios, with the former determining how oblate (``squished at the pole'') the secondary is and the latter determining how elongated it is in the equatorial plane. We found it challenging to present the dependence of the dynamical outcome on four different parameters in the same graphic. In the end, we have settled on a format in which the dependence of the dynamical outcome on the shape of the secondary is presented as a 6x6 square of color-coded tiles. Then we arrange those squares in a 6x6 pattern, representing the secondary size and initial eccentricity. Figure \ref{tiles1} shows this kind of plot for the $J_2=0.02$, $a=3 R_1$, $\theta=180^{\circ}$ slice of the set a111030 (i.e. half of the points plotted in Fig.~\ref{fig1}), while Fig.~\ref{tiles2} plots this graphic for the $J_2=0.02$, $a=7 R_1$, $\theta=180^{\circ}$ slice of the set a211030. 

The dependence on $b/c$ ratio is less pronounced, with the more oblate shapes being somewhat more prone to both chaotic rotation and barrel instability. The dependence on the equatorial elongation is more complex, with hints of structure as trends appear to reverse between the values we sampled. Overall, barrel instability is seen more often in relatively elongated secondaries. The apparent finer structure is likely due to resonances between libration frequency and the orbital motion discussed by \citet{cuk10}. Further investigations using higher resolution of initial conditions will be needed to explore this issue.

\section{The Role of Dissipation}\label{sec:diss}

Our integrations were all started with the secondary being at pericenter, with the long axis aligned with the primary, and spinning around the shortest axis (perpendicular to the orbit) at the synchronous rate. In reality, the secondary will have spin librations forced by the primary's torques \citep{md99}. If we assume that the librations are purely forced (i.e. that any free librations have been damped), the secondary will spin at sub-synchronous rate \citep{tis09, nai15a}. Assuming no spin-orbit coupling, we estimate the inertial rotation rate to be \citep{md99, tis09, nai15a}:
\begin{equation}
\dot{\phi}= n \left( 1 + {2 e \over {1 - (n/\omega_0)^2}} \right)     
\label{matt}    
\end{equation}
where
\begin{equation}
\left({\omega_0 \over n}\right)^2= {3 (B-A) \over C} = {3 (a^2 - b^2) \over a^2+b^2}
\label{omega}
\end{equation}

and $C > B > A$ are dimensionless principal moments of inertia. While it is known that spin-orbit coupling can greatly reduce forced librations in binary asteroids \citep{nai15a}, this expression is useful as it represents the end-member of libration behaviors, with the real damped rotation rate of the secondary at pericenter being in between the mean motion $n$ and $\dot{\phi}$ given by Eq. \ref{matt}.

\begin{figure}
\epsscale{.6}
\plotone{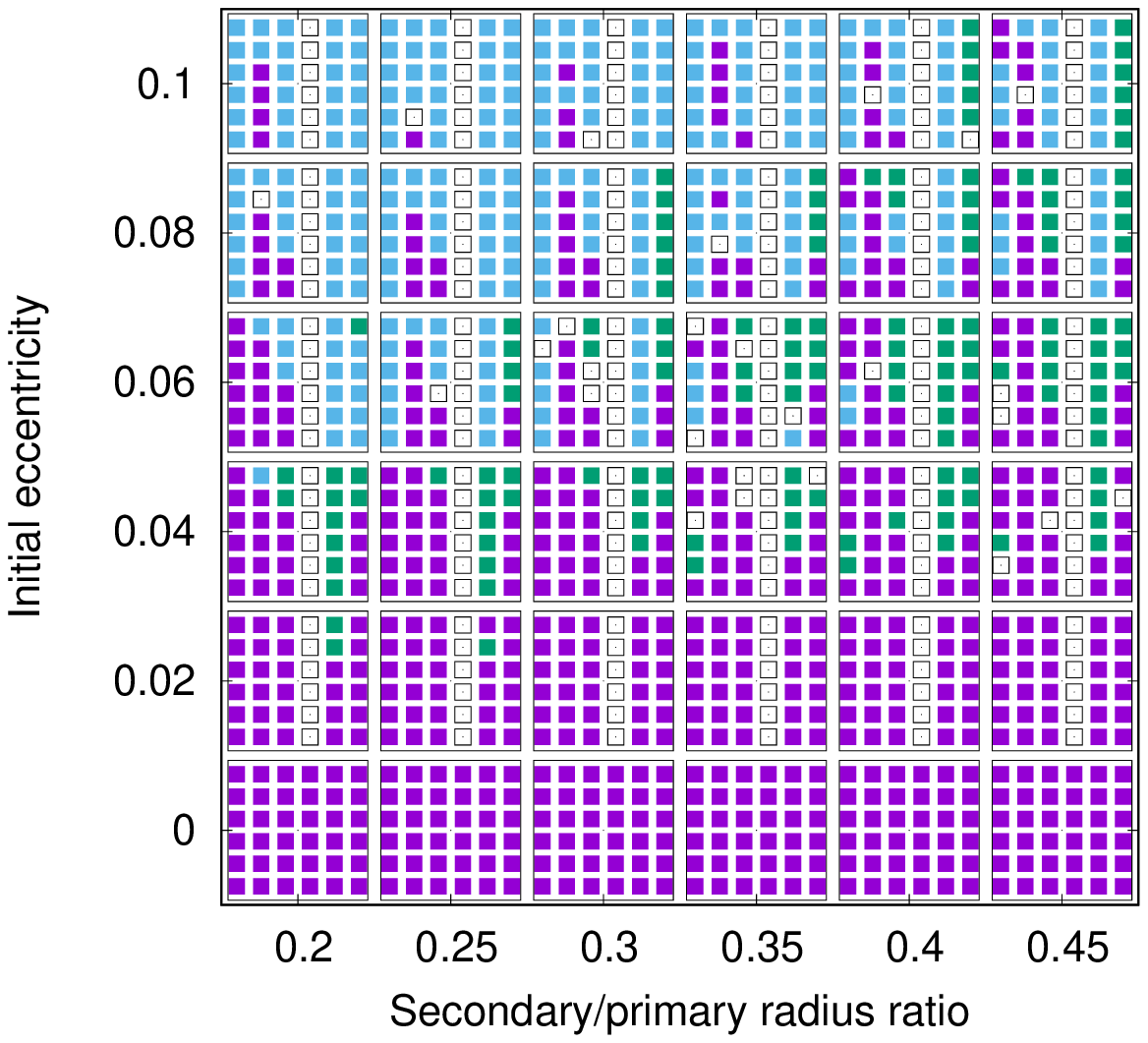}
\caption{Dependence of the dynamical outcome on the parameters of the binary, starting with approximation of no free libration. Each square point represents a simulation, and magenta symbols indicate synchronous rotation, dark green symbols indicate barrel instability, light blue symbols indicate chaotic rotation, and open symbols plot other (unclassified) outcomes. Each small 6x6 square represents dependence on the secondary's shape: the axis ratio $a/b$ changes from 1.1 to 1.6 left to right, and the axis ratio $b/c$ changes from 1.05 to 1.3 bottom to top. These small squares are then arranged according to the relevant secondary/primary radius ratio (on x axis) and initial eccentricity (on y axis). Apart from the initial rotation rate, all the system parameters are the same as in Fig. \ref{tiles1}.}
\label{ta7}
\end{figure}

In order to test the dependence of our results on the initial rotation state of the secondary, we re-ran a block of 2592 simulations with the secondary's initial spin given by Eq. \ref{matt}. All the other parameters of the were the same as in Fig. \ref{tiles1}, and the results are plotted in Fig. \ref{ta7} in the same format as in Fig. \ref{tiles1}. 

The outcomes shown in Fig. \ref{ta7} differ from those in Fig. \ref{tiles1} in several ways. The most obvious is that all non-circular cases with secondary elongation $a/b=1.4$ result in neither chaotic nor synchronous rotation. Inspection of these simulations has shown that these secondaries are in a regular retrograde rotation state. This is a direct consequence of Eq. \ref{matt}, which for $a/b=1.4$ produces $\dot{\phi}/n < 0$ for $e > 0.014$. When setting up our initial rotation we assumed that forced librations around the synchronous rotation are moderate, and that free libratons are damped. Since this value of elongation is close to the one for which $\omega_0 = n$ \citep{md99, cuk10}, libratons will always be large, and synchronous rotation may not be possible unless the orbit is very close to circular, or the spin-orbit coupling acts to reduce librations. We can conclude that for the subset of our simulations with $a/b=1.4$, assuming initially synchronous rotation is no worse than trying to estimate the libration by this simple approach.

Most of the other simulations in Fig. \ref{ta7} do not change their behavior from their counterparts in Fig. \ref{tiles1}. One moderate change is that there are a few more synchrnous outcomes on the far left-hand side of the plot, among the systems with small secondary sizes and small elongations. In these cases, Eq. \ref{matt} predicts moderate librations and simulatons assuming forced-only libration result in a few more synchronuous systems, which were largelly chaotic in Fig. \ref{tiles1}. On the upper right-hand side of Fig. \ref{ta7}, where barrel instability was more common in Fig. \ref{tiles1} among large secondaries on initially more eccentric orbits, chaotic orbits now dominate. Similarly, for the initial $e=0.06$, barrel instability is now common among large elongated secondaries, which were synchronous in our original simulations. Following \citet{nai15a}, we conclude that the spin-orbit coupling is the strongest in these systems with large secondaries, making Eq. \ref{matt} overestimate the amount of libration and there overexcite the rotation of the secondary. 

The results of the simulations shown in Fig. \ref{matt} imply that large initial free librations were not the principal cause of chaotic rotation and barrel instability in our dynamical survey. However, there is another dissipation-related question regarding barrel instability. In the limit of no spin-orbit coupling, synchronous satellites are expected to damp their free librations quickly, while the forced librations would decrease on the timescale of eccentricity damping \citep{md99, tis09}. Is it possible that barrel instability may be damped similar to free librations, rather than forced ones?
To answer this question, we ran some longer simulations in the presence of strong tidal dissipation. 

\begin{figure}
\epsscale{1.}
\plotone{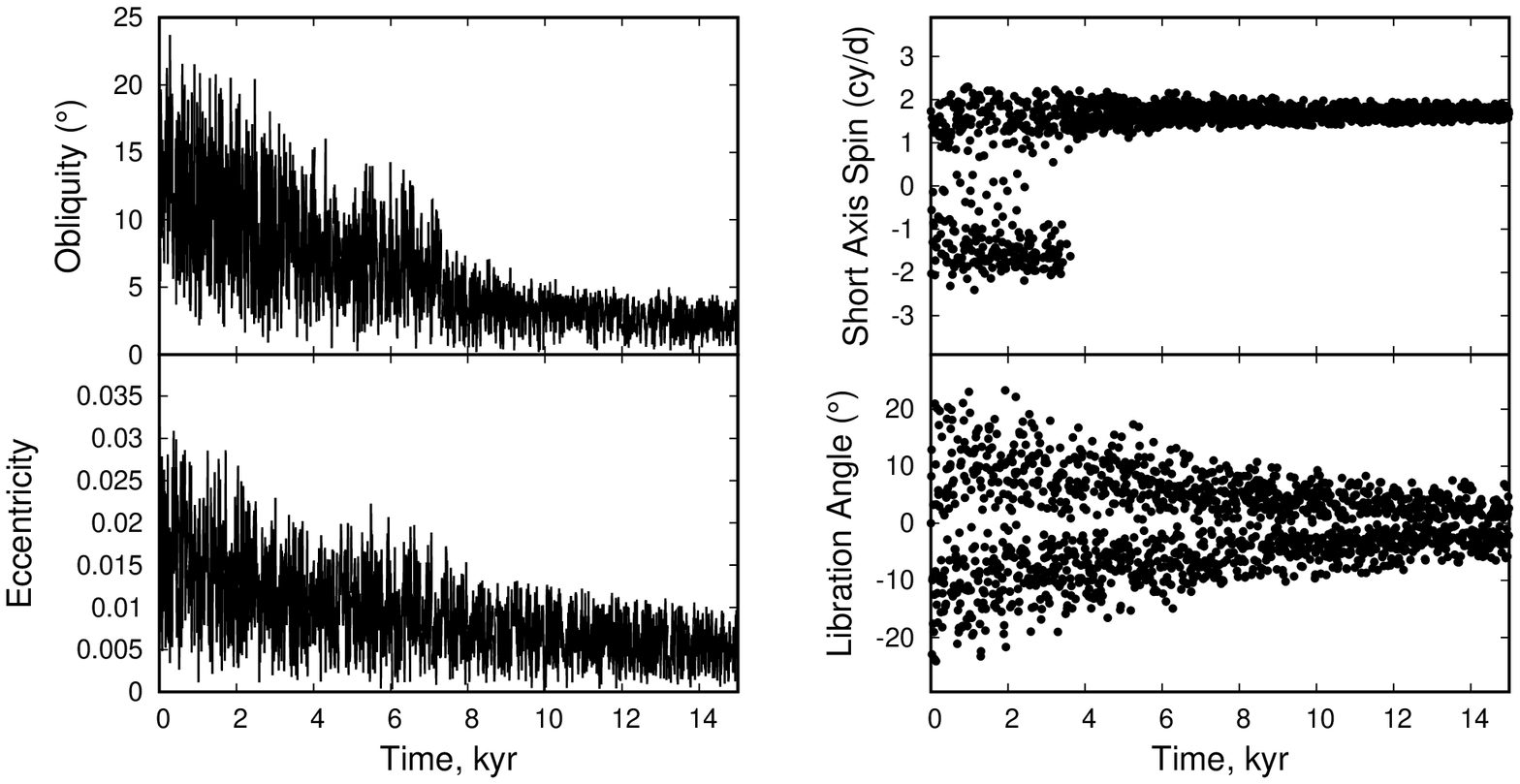}
\caption{Evolution of a binary system experiencing barrel instability in the presence of strong tidal dissipation. We used tidal parameters $Q/k_2=2.5 \times 10^6$ for the $R=$125~m secondary, comparable with the predictions of \citet{jac14}. The barrel instability ends about 4000~yr into the simulation as the eccentricity and obliquity are damped.}
\label{damp}
\end{figure}

Fig. \ref{damp} shows one such simulation using $Q/k_2=2.5e6$ for the $R=125$~m secondary (other parameters were $e_0=0.04$, $a/b=1.4$, $b/c=1.25$ for the secondary and $J_2=0.02$ for the primary).Our tidal model follows that of \citep{cuk16b}, and disspation depends only on the instantaneous rotational motion of the secondary relative to the direction of the primary. The secondary was started with synchronous rotation pericenter and was immediately undergoing barel instability. Fig. \ref{damp} shows that the damping of the librations, eccentricity and obliquity are all coupled, and that barrel instability stops (at 4000~yr) only once there has been significant eccentricity damping. Note that the secondary's obliquity and eccentricity are highly correlated, and are being damped together. When eccentricity and obliquity are free and independent, eccentricity should damp faster \citep[by a factor of seven for synchronous rotators][]{chy89}. We conclude that the secondary's rotational excitation (i.e. librations and obliquity) in this system is closely coupled to its mutual orbital eccentricity, and that the body will settle in synchronous rotation only once the system underwent significant rotational damping. In other words, chaotic rotation and barrel instablity are not short-lived phases but are likely to persist as long as the binary systems have sufficient eccentricity.

The long-term evolution of binary asteroids is outside of the scope of this paper, but we can speculate that the long-term state of the system would depend on the balance between excitation and damping. Excitation may come from interactions with the primary which in reality cannot be perfectly rotationaly symmetric \citep{jac11b}, or with semi-secular resonances with the Sun, which we did observe in our longer simulations. Tidal forces act to damp librations and eccentricities, while material transport (either between the components and within a component) can both lead to excitation and dissipation, depending on the exact mechanism. The fact that at least some secondaries are known to be eccentric and/or non-synchronous \citep{pra16} tells us that dissipation does not dominate 100\% of the time. The detailed understanding of binary asteroids long-term dynamics will require both new data and new modeling work.

\section{Implications for Real Binaries}\label{sec:real}

Here we compare our results for synthetic binaries with the known parameters of real binary systems. Since no real system is exactly situated on our initial condition grid, the comparison must be approximate. Furthermore, large uncertainties for some of the parameters of known binary systems make large swaths of our dynamical survey at least formally consistent with the properties of real pairs. Therefore all numbers provided in this section should be taken as preliminary, and this kind of rough comparison should not be considered a replacement for a dedicated dynamical study of each pair.

Since we found that the Sun is not a significant perturber on compact binaries, we simply used the simulations with the obliquity of $180^{\circ}$ from the set a111030 for these comparisons. We used one or two blocks, depending on the binary semimajor axis being close to one of our initial conditions or intermediate between two of them. In all cases we included simulations with both values of primary $J_2$ as it is usually poorly constrained. Like with binary semimajor axis, we used one or two values for the component mass ratio, depending on the correspondence between the measured value and our initial conditions. As the $b/c$ axis ratio for secondaries is usually very poorly constrained, we allowed all values for most of the binaries studied, except for 2000~DP107 (see below). We allowed the full range of $a/b$ axis ratios allowed by the uncertainties, and did not attempt to weight the probabilities within this range; we find that narrowing the range did not qualitatively change the outcome so we used the simplest approach where every allowed synthetic binary was counted the same.

The binary eccentricity is maybe the most important parameter that determines the dynamics of the binary. For real binaries the eccentricity is often poorly constrained, leading to a wide variety of allowed outcomes. For the sake of simplicity, we decided to compare the observed eccentricities for real binaries to the {\it averaged} eccentricities in our simulations. While this opens a possibility of missing some unlikely cases, such as very low eccentricity in a system being a chance instantaneous value within a wide oscillation, it also makes comparisons easier and faster. Furthermore, since our comparisons are only probabilistic, a much more complex analysis would likely only slightly change the final probabilities.

The results of our comparison between synthetic binaries and real systems are presented in Table \ref{real}. For each of the eight well-known binary NEA systems we present the observed system parameters and the relevant reference. Then we give the breakdown of dynamical outcomes for all synthetic binaries that are consistent with the real binary given the uncertainties in the observed parameters. The numbers for the synchronous rotation, barrel instability and chaotic rotation may not add up to 100\% due to the presence of ``other'' outcomes (usually transitions between the other three). The numbers should be taken critically, especially for binaries for which their eccentricity and/or shape are poorly constrained and a large swaths of our dynamical survey are being sampled. The most meaningful number is zero, in the sense that if some outcome is listed as $0\%$ for a binary, we can exclude it from consideration. 

\begin{table}
\centering
\caption{Parameters of the real binary systems discussed in this Section. The last three columns are the percentages of synthetic binaries consistent with the observed system that are in synchronous rotation, barrel instability or chaotic rotation, respectively. The abbreviation for references are as follows: NM15, \citet{nai15b}; N+18, \citet{nai18}; N+20, \citet{nai20}; S+15, \citet{sch15}; S+21, \citet{sch19}; P+16, \citet{pra16}.\label{real}}
\begin{tabular}{lcccclccc}
\hline
Designation & $a/R_1$ & $e$ & $R_2/R_1$ & $a_2/b_2$ & Reference & Synch. \% & B.I. \% & Chaotic \%\\
\hline
(66391) Moshup & 3.4 $\pm$ 0.4 & $<$0.006 & 0.42 $\pm$ 0.03 & 1.3$^{+0.3}_{-0.1}$ & S+21 & 100 & 0 & 0\\
(88710) 2001~SL9 & 3.5  $\pm$ 0.6 & $\le$ 0.07 & 0.24 $\pm$ 0.02 & $\le$1.2 & S+21 & 59 & 0 & 38\\
(175706) 1996~FG3 & $3^{+0.6}_{-0.4}$ & $\le$ 0.07 & 0.29 $\pm$ 0.02 & 1.3 $\pm$ 0.2 & S+15 & 62 & 15 & 20\\
(65803) Didymos & 3 $\pm$ 0.1 & $<$0.05 & 0.19 $\pm$ 0.04 & unknown & N+20 & 47 & 12 & 40\\
(185851) 2000~DP107 & 6.2 $\pm$ 0.3 & 0.019 $\pm$ 0.01 & 0.37 $\pm$ 0.03 & 1.2 $\pm$ 0.1 & NM15 & 69 & 3 & 21\\
(35107) 1991~VH & 5.4 & 0.05 & 0.44 & high & N+18 & 5 & 0 & 88\\
(7088) Ishtar & 4.4 & $<$0.18 & 0.42 & 1.5 & P+16 & 53 & 37 & 8\\
(66063) 1998~RO1 & 3.6 & $<$0.06 & 0.48 & 1.5 & P+16 & 63 & 37 & 0\\
\hline
\end{tabular}
\end{table}

{\bf Moshup} (formerly 1999~KW4) is one of the best characterized binaries, and has served as the prototype for the whole small binary population \citep{ost06, sch06, fah08}. \citet{sch19} have recently measured its orbital drift, and found a slow outward drift, which could be attributed to either BYORP or tides. BYORP drift predicted on the basis of the shape model is also outward, but was expected to be faster \citep{mcm10b}. Radar observations and subsequent analysis restrict Moshup to very low binary eccentricities, and all synthetic binaries in the same part of the phase space (a few dozen) were in a synchronous dynamical state. Of course, the same observations and analysis of Moshup that measured the eccentricity also reported that the secondary was in synchronous rotation, making our result unsurprising. Still, this was a valuable check on our approach and the results are encouraging.

{\bf 2001~SL9} is another binary for which \citet{sch19} reported a non-zero orbital drift, this time inward, possibly indicating the BYORP effect. As the eccentricity of this pair is poorly constrained, we get several hundred synthetic binaries consistent with this pair. The majority of our synthetic analogues have synchronous secondaries, but a large minority are chaotic, with barrel instability being absent. Relatively modest elongation of the secondary is probably crucial in making the synchronous rotation the most common dynamical outcome among the corresponding synthetic binaries. If the BYORP detection holds, the secondary must be synchronous, and the eccentricity is probably much below the allowed upper limit of $e_{max}=0.07$.  

{\bf 1996~FG3} is a binary for which \citet{sch15} found an orbital drift consistent with zero, interpreted as an equilibrium between BYORP and tides. As the eccentricity of 1996~FG3 is poorly constrained, a wide range of initial conditions in our survey are within the allowed parameter space, and hundreds of synthetic binaries have been identified as possible matches. Just about half of secondaries are in synchronous rotation, about a third in chaotic, and about 15\% are in barrel instability. This breakdown probably has more to do with how widespread these outcomes are in the whole of parameter space than specifically for 1996~FG3, and the safest conclusion would be that all three outcomes are possible. \citet{sch15} reported not seeing any signs of chaotic rotation, but it is not clear if this would also exclude barrel instability which was not recognized before as a distinct dynamical state (and may less readily detectable through light curves). 1996~FG3 will be one of the targets of the Janus mission \citep{janus}.

{\bf Didymos} is the target of the DART mission, which will alter the orbit of the secondary by using a kinetic impactor \citep{che18}. The published properties of the binary are not all well constrained, and a wide range of synthetic systems are consistent with observations. Among them we get all three dynamical outcomes, without a majority. The relatively small size of the Didymos secondary makes chaotic rotation more likely than it would otherwise be, but this is partially offset by the very small binary separation. \citet{agr21} have recently found that Didymos B may be vulnerable to the barrel instability, and that this behavior may be triggered by the impact. 

{\bf 2000~DP107} is another very well-characterized binary NEA. It is known to have a low eccentricity, and we were also able to restrict the shorter axis ratio to $1.1 \le b/c \le 1.25$. Only a couple of dozen synthetic secondaries fall within the range of parameters allowed for 2000~DP107, and overwhelming majority are synchronous (with barrel instability ruled out). We conclude that the comparison between 2000~DP107 and our simulation suggests that its secondary is likely to be synchronous, in line with other available data on this system \citep{nai15b}.

{\bf 1991~VH} is the most prominent binary NEA known to be in chaotic rotation \citep{nai15a, nai18}. While the exact parameters of its shape are not known, we restricted ourselves to elongated cases $a/b \ge 1.5$, and allowed for average eccentricities in the $0.04 \le e \le 0.06$ range (in the absence of formal uncertainty). An overwhelming majority of the dozen synthetic pairs consistent with these criteria were in chaotic rotation, encouraging us that our results are relevant to real binaries. 1991~VH is another target of the Janus mission \citep{janus}

{\bf Ishtar} is considered a future target for BYORP measurements. While the orbital eccentricity is poorly constrained, the secondary/primary ratio and secondary elongation are relatively high, making both synchronous rotation and barrel instability likely rotation states at this point.

{\bf 1999~RO1} is another likely candidate for future BYORP measurements. Its parameters are overall similar to those of Ishtar, and likewise both synchronous rotation and barrel instability are reasonably likely on the basis of comparison with our simulations.

The above cases are directly comparable to our integrations. In future work we hope to address smaller secondaries, more separated and more eccentric systems, and also equal-size pairs. In particular, outer satellites in triples like 1999~CC \citep{bro11}and 2001~SN263 \citep{bec15} are known to be asynchronous, but their parameters are outside the our box of initial conditions.

\section{Conclusions}\label{sec:con}

Here we used numerical integrations of a large number of synthetic binary asteroids to explore the importance of orbital and shape parameters on the dynamical state of the binary. Our conclusions are as follows:

1. We identify a new rotational state that is possible for elongated secondaries, and call it the ``barrel instability''. In this state the secondary is in non-principal axis rotation in many ways similar to fully chaotic rotation, but its longest axis stays approximately aligned with the direction to the primary. 

2. A large majority of binary asteroids with near-synchronous secondaries settle into one of the three basic dynamical states, in which the satellites are respectively in synchronous rotation, chaotic rotation, and barrel instability.

3. Heliocentric orbital elements, tilts between heliocentric and mutual orbits, and the oblateness of the primary do not have major effects on the spin-orbit dynamics of the secondaries in close binaries.

4. Binary eccentricity is the major driver of chaotic rotation and barrel instability, but due to spin-orbit interactions the three dynamical outcomes are not clearly separated in eccentricity.

5. Another crucial parameter is binary separation in terms of {\it secondary} radius. Smaller and/or more distant secondaries are more prone to chaotic rotation, while larger/closer secondaries are more often synchronous or experience barrel instability.

6. Elongated secondaries are more susceptible to barrel instability.

7. When we compare our results to the known binaries, our simulations correctly identify the companion of Moshup as synchronous, that of 2000~DP107 as likely synchronous, and the moon of 1991~VH as a very likely chaotic rotator.

8. Our results allow for the possibility that the secondary of 1996~FG3 is undergoing barrel instability. Positive determination of 1996~FG3 rotation state would either strengthen the BYORP-tide equilibrium paradigm if it is synchronous, or falsify the assumption of the BYORP-tide equilibrium if this binary is found to be in barrel instability.

\begin{acknowledgments}
The authors wish to thank two anonymous reviewers whose comments greatly improved the paper. M\'C is supported by NASA Solar System Workings program award 80NSSC21K0145.
KJW was supported in part through the NASA Solar System Exploration Research Virtual Institute node Project ESPRESSO, cooperative agreement number 80ARC0M0008.
\end{acknowledgments}

\bibliography{barrel_refs}{}
\bibliographystyle{aasjournal}



\end{document}